# A study of seismology as a dynamic, distributed area of scientific research

Caroline S. Wagner[1] and Loet Leydesdorff
University of Amsterdam
April 2003

**Abstract**

Seismology has several features that suggest it is a highly internationalized field: the subject matter is global, the tools used to analyse seismic waves are dependent upon information technologies, and governments are interested in funding cooperative research. We explore whether an emerging field like seismology has a more internationalised structure than the older, related field of geophysics. Using aggregated journal-journal citations, we first show that, within the citing environment, seismology emerged from within geophysics as its own field in the 1990s. The bibliographic analysis, however, does not show that seismology is more internationalised than geophysics: in 2000, seismology had a lower percentage of all articles co-authored on an international basis. Nevertheless, social network analysis shows that the core group of cooperating countries within seismology is proportionately larger and more distributed than that within geophysics. While the latter exhibits an established network with a hierarchy, the formation of a field in terms of new partnership relations is ongoing in seismology.

## *1. Introduction*

A number of studies have demonstrated that international linkages in science and technology are increasing. (Glänzel 2001, others) The linkages can be observed at the global level (Wagner and Leydesdorff, 2003) and at the level of scientific disciplines. Linkages within disciplines or fields of science can be observed by exposing the networks of scientific citations and co-authorship relations that are created as scientists cite each other's work within scientific articles or acknowledge each other as co-authors (Mullins, 1988; Wouters, 1999). When examined over time, these citation and co-authorship networks enable us to reveal the dynamics of emerging fields.

In this study, we use this method to test the hypothesis that emerging fields of science are more likely than older established fields to be internationally networked. This is based in part on the expectation that emerging fields are also likely to have features reflecting a "Mode 2" method of operating: they are interdisciplinary, team-based research. (Gibbons et al., 1994) Moreover, they are more likely to incorporate new information and communication tools into their operations, and thus be more likely to be networked.

In order to test this hypothesis, we examined a field of science that has several features suggesting that it is both dynamic (evolving rapidly) and highly internationalised (many co-authorships across countries). Seismology – the study and analysis of seismic waves within the Earth's mantel – provides a good case study for testing this hypothesis. First, seismology is a field that has emerged largely because of the capabilities offered by new technologies: seismometry is highly technical and the complexity of the resulting data requires computer analysis. Second, the research requires access to data collected from and shared around the globe. Third, several countries have invested in unique research equipment and resources that can only be made available to researchers from other countries through collaboration.[2] Fourth, many governments are interested in funding research that may help anticipate, if not an earthquake itself, then at least the nature of damages that may result. Fifth, the research does not have direct commercial application, suggesting that most of the collaborative research will be published in open literature.

Many analysts have noted that the ability of scientists in different countries to collaborate has increased significantly over the past 15 years, with improving access to the Internet being only one of these reasons. (Gibbons *et al.*, 1994 and others) The willingness of governmental and non-governmental organizations to provide technical assistance to other nations, including exchange of experts, when a major earthquake takes place has also increased the extent to which seismologists and other earth scientists meet and share research interests with each other. This is particularly important for the decade we studied, 1990 to 2000. In 1994, California experienced the Northridge earthquake. One year later, in 1995, Japan experienced the Kobe earthquake. Both of these events led to exchanges of scientists and engineers, and both earthquakes spurred new research into seismology, structural engineering, and earthquake sciences.

---

[1] Correspondence to: C. Wagner, Amsterdam School of Communications Research, University of Amsterdam, Kloveniersburgwal 48 1012 DX, Amsterdam, Netherlands, +31-20-525-7257, wagner@pscw.uva.nl.

[2] For example, Japan has invested in a shake table that simulates earthquake movement and provides valuable data on the possible affects of tremors on engineered structures. Russia has dug a bore hole in a seismologically active area to enable testing of different features of geology. The United States has invested in a global seismographic network that provides continual readings of seismological activities at dozens of locations around the globe.

## 2. Methodology

In order to explore the structure of the field of seismology and to test whether it shows international dynamism, we conducted a series of cascading analyses. The first part of the analysis identified the citing relationships to uncover the cluster of journals that scientists identify when citing other published work relevant to their article. Secondly, within the decade of 1990 to 2000, we collected all addresses of the articles in the relevant journal clusters for 3 different years (1990, 1994, and 2000). We then used this in the third step to detail networks of linkages among scientists at the national level. Finally, in the fourth step, we analysed the percentages of international co-authorships within these years to view the extent to which seismology had grown as an international discipline. Each step is described in more detail below.

### Defining the field

Research shows that scientific journal-journal citations can be exploited to expose patterns of interrelationships within and among fields of science. (Carpenter & Narin, 1975; Doreian & Fararo, 1985); Tijssen et al., 1987). Leydesdorff and Cozzens (1993) suggest that disciplines of science can be operationalised in terms of *journal sets*. As such, journal-journal citations can be used to track changes in the disciplinary structure of science. The patterns provide a method for studying the structure of a field in a single year as well as over time. The citation relationship among journals reveals a structure of the literature that scientists view as relevant to their work. An analysis of these citation relationships reveals "clusters" that can be visually depicted. This method of journal-journal citation mapping is applied here to see if the structure of citations related to the field of seismology has changed, how is it related to other fields, and whether the field has been influenced by international linkages.

Using the method developed by Leydesdorff and Cozzens (1993) and later applications by Leydesdorff, Cozzens, and Van den Besselaar (1994), we identify the journals most closely associated with the field. In describing the methodology, Leydesdorff and co-authors note out that journal-journal citation relations "contain information about field and subfield structures at a sufficient level of aggregation for the construction of indicators…" (Leydesdorff et al. 1994) Journal publications are not stable over time: The incremental change can become an indicator in itself by revealing the change of structure at the level of the field over time. For example, the inclusion of new journals within a cluster of related journals may indicate the extension or further differentiation of the cluster or it may indicate qualitatively new developments in the field. The emergence of new clusters, or the merging of clusters, can also be indicative of structural changes in a field.

The Leydesdorff-Cozzens method enables the creation of journal-journal citation maps based on factor or *eigenvector* analysis of the citations of journals in the *Science Citation Index* (SCI) and the Journal Citation Reports. The data we report here are drawn from the databases of the Institute for Scientific Information *SCI* and JCR CD-Rom data. The latter data consist of listings of all journals processed and included on the CD-Rom for 1994-2001. Co-author names and affiliations were obtained from the Science Citation Index extended available through the Web of Science in May 2002.

## Revealing the structure of the field

The first step in identifying initial journals to work with involved using the Science Citation Index to identify journals with titles related to the field of seismology. We searched for the words "earthquake" and "seismology" in journals available in March 2002 on the Web of Science. A number of journals were initially considered, although two journals had names that appeared directly relevant: *The Journal of Seismology* and *The Bulletin of the Seismological Society of America*. Using a factor analysis, we found that, in 2000, the *Bulletin of the Seismological Society of America (BSSA)* is the journal most closely associated with the field of seismology. BSSA served as the central tendency journal (CTJ) for the analysis. Leydesdorff and Cozzens (1993) have proposed "central tendency journals" as yardsticks for measuring structural change in fields of science over time. Central tendency journals are defined as seed journals that exhibit the highest correlation with the eigenvector that represents the cluster at the network level. Central tendency journals exhibit more stability than journals that are less central to the cluster. We examined the citing patterns of the BSSA in order to identify the action parameter associated with the field. Journals being cited by articles in BSSA indicate the structure of the field as viewed by the authors in 2000.

The cluster of journals that emerge from the citing patterns around BSSA in 2000 reveal the structure of the field in that year. Table 1 displays two fields of science closely related to seismology. The first box highlights those journals loading on factor 1, indicating the field of geophysics. The solid box reveals BSSA as the "central tendency journal" for 2000 as well as its citing galaxy, loading on factor 2.

**Table 1: A factor analysis of the citing patterns of journals related to seismology in 2000**

| **Journal** | Factor 1 | Factor 2 | Factor 3 | Factor 4 |
|---|---|---|---|---|
| *Earth, Plants and Space* | *.96755* | .13346 | .03939 | .00451 |
| *Journal of Geophysics Research* | *.92073* | .09846 | .11211 | -.04233 |
| *Geophysics Research Letters* | *.89984* | .08973 | .19592 | -.05187 |
| *Geophysics Journal International* | *.86183* | .25644 | .00527 | .16200 |
| *Physics of Earth and Planetary Interiors* | *.82490* | .05293 | .08132 | .08248 |
| *Pure and Applied Geophysics* | *.82072* | .47634 | .03008 | .06451 |
| *Tectonophysics* | *.74253* | .15776 | .04004 | .09377 |
| **Bulletin of the Seismological Society of America** | .29372 | **.92325** | .04560 | -.07577 |
| **Soil Dynamics and Earthquake Engineering** | -.18017 | **.89607** | -.07031 | -.23901 |
| **Journal of Seismology** | .38826 | **.84096** | -.03825 | -.09281 |
| **Natural Hazards** | .60390 | **.71365** | -.01758 | -.02778 |
| **Annali di Geofisica** | .59009 | **.60293** | -.19081 | .19119 |
| **Engineering Geology** | .11106 | **.53651** | .01291 | .06764 |
| Science | .08075 | -.06509 | .93182 | .00392 |
| Nature | .03365 | -.08399 | .93085 | .00752 |
| Current Science | .12658 | .08081 | .86149 | -.02022 |
| Geophysics | -.01183 | .06800 | -.15294 | .80088 |
| Earthquake Engineering and Structural Dynamics | -.16112 | .24468 | .18170 | -.62480 |

This factor analysis provides two key pieces of information for this case, shown in Table 1. One is that the field of geophysics, represented in the first cluster of variables loading on Factor 1 (enclosed in a dotted box), is tightly clustered and well defined. The second finding is that BSSA is the central tendency journal for the field of seismology loading on Factor 2 (enclosed in a solid box)[3]; BSSA has positively correlated citation patterns with five other journals within the factor structure:

- *Soil Dynamics and Earthquake Sciences*,
- *Journal of Seismology*,
- *Natural Hazards*,
- *Annali di Geofisica*, and
- *Engineering Geology*.

The factor analysis for the citing relationships results in a plot of the stimulus space showing a 2-dimensional drawing of the citation relationships. The next section displays and discusses the figures that emerge from this analysis.

## Mapping the international network of the field

In order to test our thesis, for each of the clusters identified in the factor analysis, "geophysics" and "seismology," we calculated the percentages of international co-authorship as a share of all papers published in that year for that cluster. Then, we mapped the networks of both fields in 2000, and the combined field in 1990. This is done using the following techniques:

1. For each of the cluster of journals, we collected all the related records from the Web of Science for that year. We saved these with authors and addresses so that we could identify linkages among authors by country.
2. Using custom computer programs written for these purposes by one of the authors, we sorted the author names and addresses into files that allowed us to count international co-authorships.
3. Taking the address file created as part of this analysis, we imported this into UciNet to create an affiliations file (Borgatti et al., 2002), and then this data is exported to Pajek network analysis software to draw a social network.[4] A core analysis conducted on this network reveals the most intense relationships among collaborating countries in both years.
4. In order to normalize for the size effects among countries, we also used the cosine between the vectors for individual countries as input to the visualization using Pajek network analysis software.(Hamers et al., 1989) This analysis provides us with a means to explore and compare the hierarchy in the networks (Wagner & Leydesdorff, forthcoming).

---

[3] Performing the same factor analysis on the Journal of Seismology shows that the Bulletin of the Seismological Society of America is also revealed to be the central tendency journal.
[4] The program Pajek is freely available for non-commercial use at http://vlado.fmf.uni-lj.si/pub/networks/pajek

Then we compare the two clusters. This is done in an effort to see whether and how seismology is more heavily internationalised than geophysics.

The steps detailed above reveal the structure of the field in 2000. The next question we asked is how the field has changed over time. Using the structure of the field in 2000, and maintaining the BSSA as the central tendency journal, we reconstructed the field historically for years 1998, 1996, and 1994, with a network comparison for 1990. The concept of *extending back* historically rather than beginning in the early 1990s and working forward is done so that we can examine the evolution of the field over time based on its construction in 2000. If one systematically accounts for delineation in the groupings ahead of time, one risks making a prediction of performance with reference to an outdated unit. The point is to identify the dynamics of the field: The validity of this method is demonstrated graphically by the fact that the journal *Seismology* was only initiated in 1999. If we had fixed the journal set only to those available in 1994, we would have missed the entrance of the new journal. So we began with the structure as it is revealed, and reconstruct the field's development from the perspective of citing relationships in 2000.

For the purposes of this case study, the set of *citing* journals for each year is used. The central tendency analysis is used to show local densities in the overall network for each year. Clusters can be compared over years, even though they are not necessarily tied to the same central journal. This helps illustrate the evolution of the field over time. The evolutionary illustrations show the relations within the cluster over the years by showing the relevant citation environment. The change in the data set can be used as an indicator of restructuring within the field.

## *3. Findings: The Emergence of Seismology from within the field of Geophysics*

Our analysis shows that seismology emerges out of geophysics as an independent field of science in terms of its citing environment over the decade of the 1990s. It did not exist as an independent field in the early part of the decade. This is illustrated in the four figures displayed below. The figures display the citing galaxy, based on factor analysis, of the journals in that year. The evolution of the field is illustrated by the emergence of a citing environment for seismology, but one that does not take shape until 1996, and even then, is not stable and receded in 1998, and then re-emerges in 2000. Figure 1 illustrates and reveals the shape of the field of seismology, compared to geophysics, in 2000. Figures 2, 3, and 4 illustrate the evolution of shape of the field of geophysics from 1994, and the emergence of seismology as a separate field over time.[5]

[5] In the uneven year 1995, the *Bulletin of Seismological Society of America* exhibits interfactorial complexity between geophysics and a factor that otherwise consists only of *Earthquake Engineering and Structural Dynamics*. In the period 1996-2001 an independent second factor can be distinguished, but not in 1998. *BBAS* functioned as a central tendency journal in 1996, 1997, 2000, and 2001, but not in the years 1993, 1995, 1998, and 1999. In 2001, the seismology group was further extended with the journals *Pure and Applied Geophysics* and the *Journal of Computer Acoustics*. *Natural Hazards*, however, was in this year no longer part of the citation environment of *BBAS* at the one percent level.

**Figure 1. Clusters of journals relating to seismology based on citing patterns, 2000**

PLOT OF STIMULUS SPACE

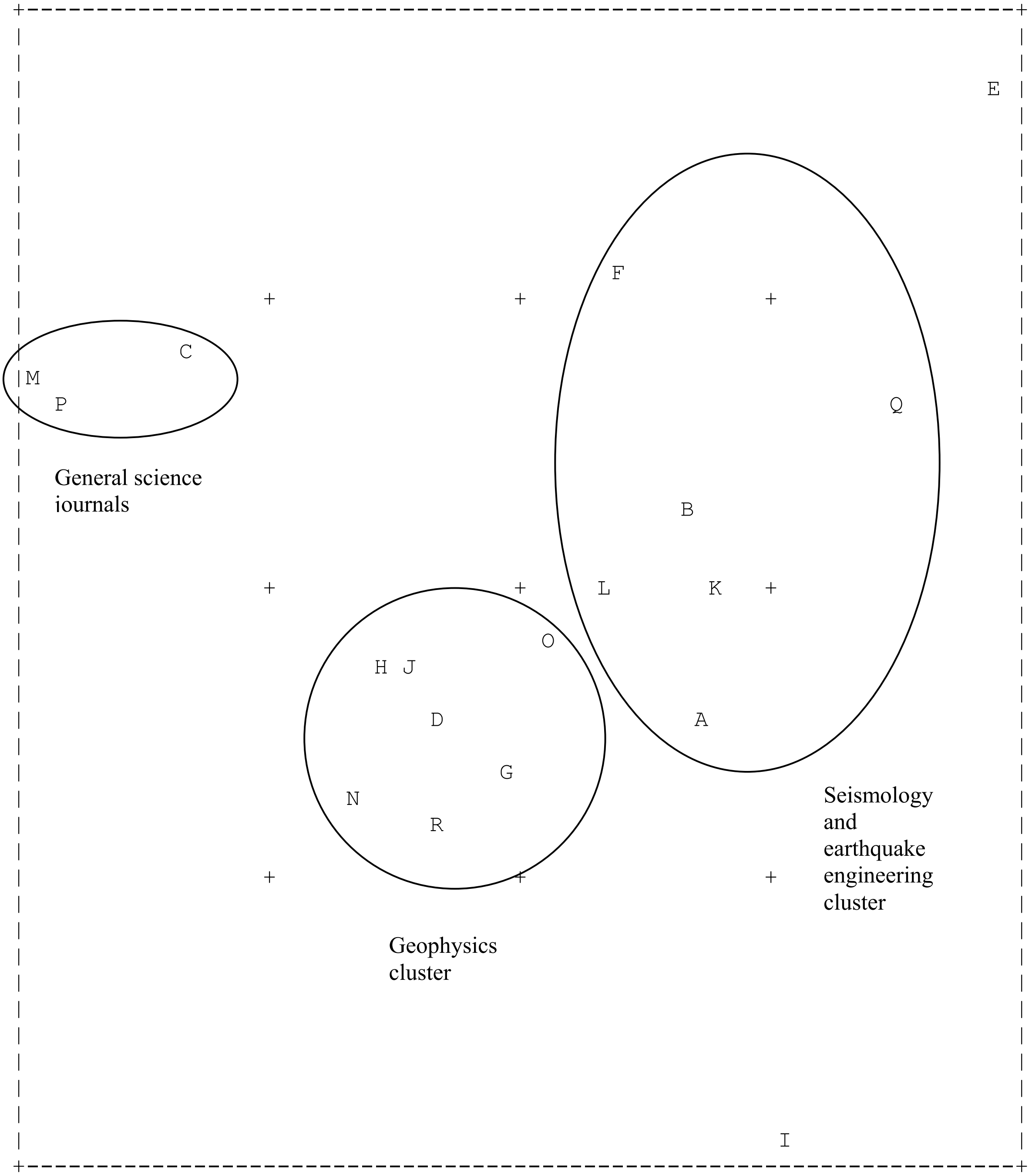

X = DIMENSION 1 Y = DIMENSION 2

Legend:

| Indicating letter | **Seismology Cluster Journal Names** |
|---|---|
| A | Annali di Geofisica |
| B | Bulletin of the Seismological Society of America |
| F | Engineering Geology |
| K | Journal of Seismology |
| L | Natural Hazards |
| Q | Soil Dynamics and Earthquake Engineering |
| | **Geophysics Cluster Journal Names** |
| D | Earth Planet and Space |
| G | Geophysics Journal International |
| H | Geophysics Research Letters |
| J | Journal of Geophysics Research |
| N | Physics of Earth and Planetary Interiors |
| O | Pure and Applied Geophysics |
| R | Tectonophysics |
| | |
| | **General Science Cluster Journal names** |
| C | Current Science |
| P | Science |
| M | Nature |
| | |
| | **Other Related but not Clustered Journals** |
| E | Earthquake Engineering and Structural Dynamics |
| I | Geophysics |

By examining the figures, one can see that in 1994, illustrated in Figure 2, three clusters are evident. These are:

1) The geophysics cluster, containing the *Bulletin of Seismological Society of America*, the *Physics of the Earth and Planetary Interiors*, *Geophysical Journal International*, *Journal of Geophysical Research-Solid Earth*, and the *Geophysical Research Letters*.
2) A general sciences cluster containing *Nature*, *Science*, and *Current Science India*.
3) A tectonics and geology cluster containing *Tectonics*, *Tectonophysics*, and the *Geological Society Of America Bulletin*.
4) Also in the citing environment, but not in the citing cluster are: *Geophysics*, the *Journal of the Acoustical Society of America*, and *Earthquake Engineering & Structural Dynamics*.

In 1996, illustrated in Figure 3, three clusters remain evident, but their constitution is different. The field of tectonics has been subsumed into geophysics, and seismology and earthquake engineering emerges as a new cluster. The clusters now represent:

1) The geophysics cluster, which now appears *without* the *Bulletin of the Seismological Society of America.* The other journals evident in 1994 remain the same, but they are joined by: *Pure and Applied Geophysics*, and *Tectonophysics*.
2) The seismology and earthquake engineering cluster includes the *Bulletin of the Seismological Society of America, Earthquake Engineering & Structural Dynamics,* and, *Soil Dynamics and Earthquake Engineering.*
3) The general sciences cluster with *Nature*, *Science*, and the *Proceedings of the National Academy of Sciences* (USA).

In 1998, illustrated in Figure 4, seismology has rejoined geophysics. Tectonics also remains as part of the geophysics cluster. Earthquake engineering remains a separate cluster. The geophysics cluster has been joined by the *Journal of Physics of the Earth* (item H). The general sciences cluster resembles that of 1994.

To return to 2000, illustrated in Figure 1, it is possible to see a significant change in the structure of the field between 1994 and 2000. Geophysics (with tectonics) remains a tight cluster with no new entrants. Seismology, however, exhibits a considerable change. Earthquake sciences and seismology have merged into a single cluster with *Soil Dynamics* appearing much closer to seismology than before. In addition, there are three new entrants to the seismology cluster: *Journal of Seismology*, *Natural Hazards*, and *Annali di Geofisica*.

This suggests that the field took on an increased self-definition over time, strengthened by the entrance of new journals into the field, ones that quickly joined the citing cluster.

**Figure 2: Clusters of journals relating to seismology based on citing patterns, 1994**

PLOT OF STIMULUS SPACE

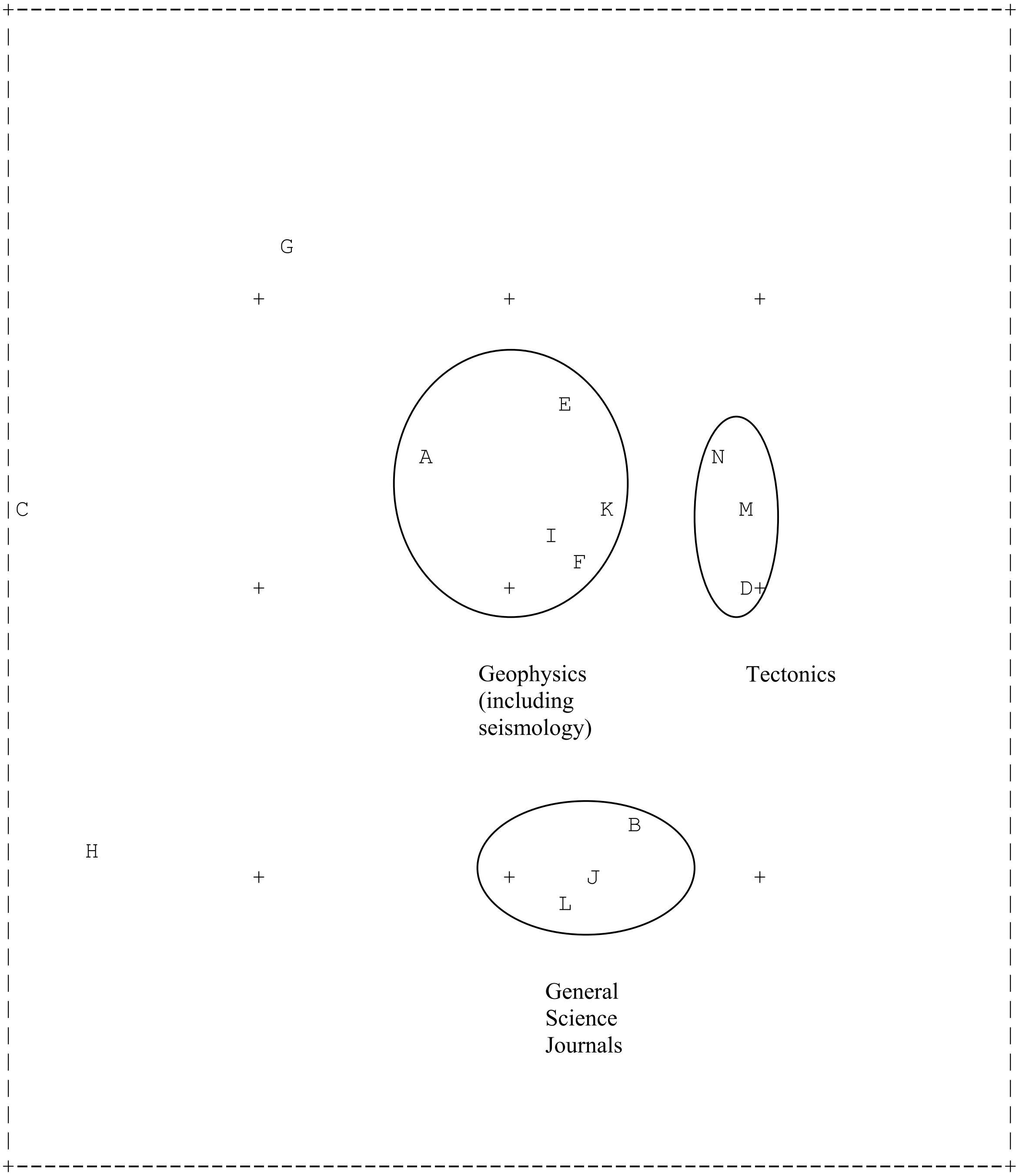

X = DIMENSION 1     Y = DIMENSION 2

Legend:

| Indicating letter | **Geophysics-Seismology Cluster Journal Names** |
|---|---|
| A | Bulletin of the Seismological Society of America |
| E | Geophysics Journal International |
| F | Geophysics Research Letters |
| I | Journal of Geophysics Research |
| K | Physics of Earth and Planetary Interiors |
| | |

| | **Tectonics Cluster Journal Names** |
|---|---|
| D | Geological Society of America Bulletin |
| M | Tectonics |
| N | Tectonophysics |
| | |
| | **General Science Cluster Journal names** |
| B | Current Science |
| J | Nature |
| L | Science |
| | |
| | **Other Related but not Clustered Journals** |
| C | Earthquake Engineering and Structural Dynamics |
| G | Geophysics |
| H | Journal of the Acoustical Society of America |

**Figure 3: Clusters of journals relating to seismology based on citing patterns, 1996**

PLOT OF STIMULUS SPACE

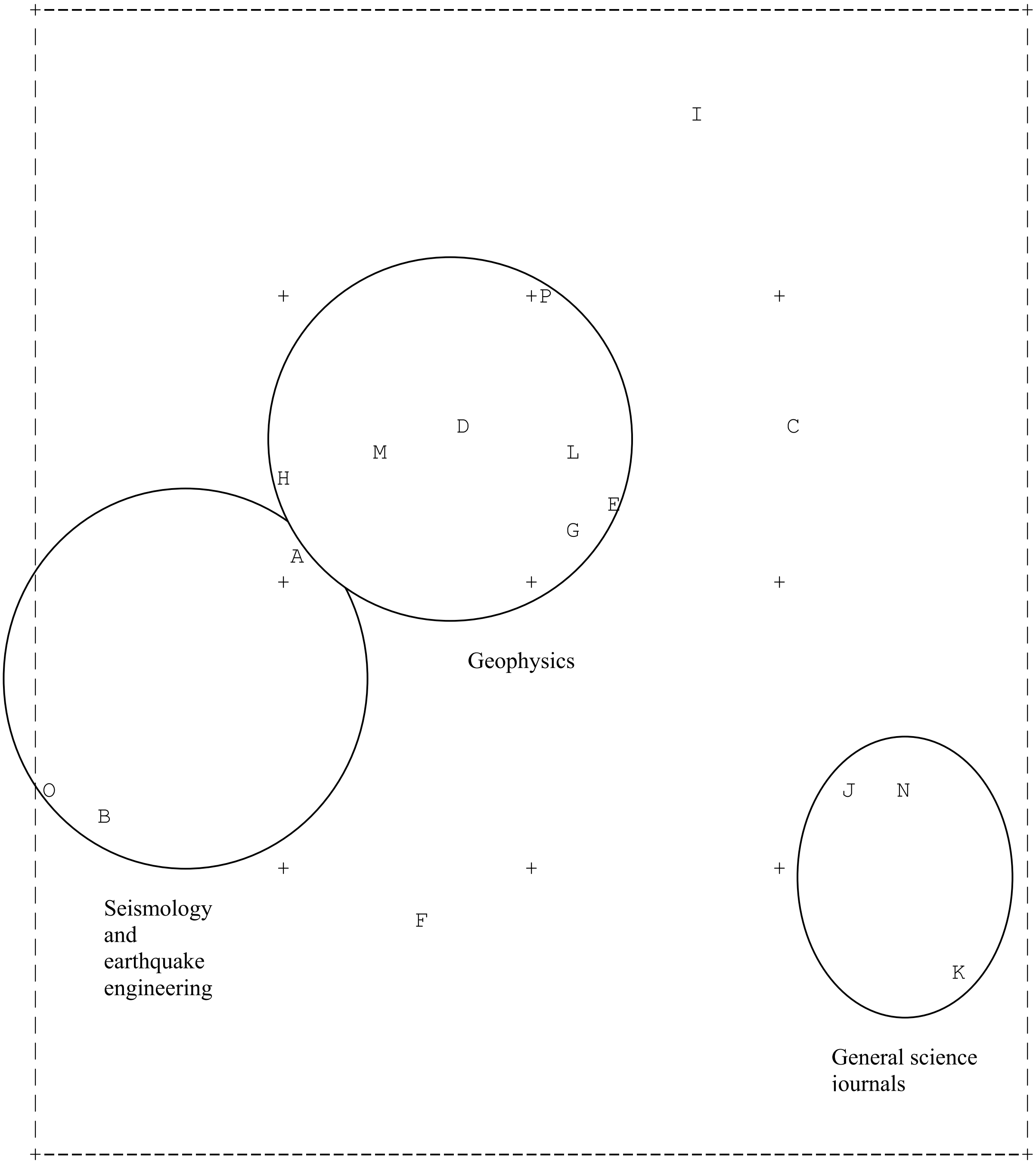

X = DIMENSION 1 Y = DIMENSION 2

Legend:

| Indicating letter | **Seismology Cluster Journal Names** |
|---|---|
| A | Bulletin of the Seismological Society of America |
| B | Earthquake Engineering and Structural Dynamics |
| O | Soil Dynamics and Earthquake Engineering |
| | |
| | **Geophysics Cluster Journal Names** |
| D | Geophysics Journal International |
| E | Geophysics Research Letters |
| G | Journal of Geophysics Research |
| H | Journal of Physics of the Earth |
| L | Physics of Earth and Planetary Interiors |
| M | Pure and Applied Geophysics |
| P | Tectonophysics |
| | |
| | **General Science Cluster Journal names** |
| J | Nature |
| K | Proceedings of the National Academy of Sciences |
| N | Science |
| | |
| | **Other Related but not Clustered Journals** |
| C | Geological Society of America Bulletin |
| F | Geophysics |
| H | Journal of Structural Geology |

**Figure 4: Clusters of journals relating to seismology based on citing patterns, 1998**

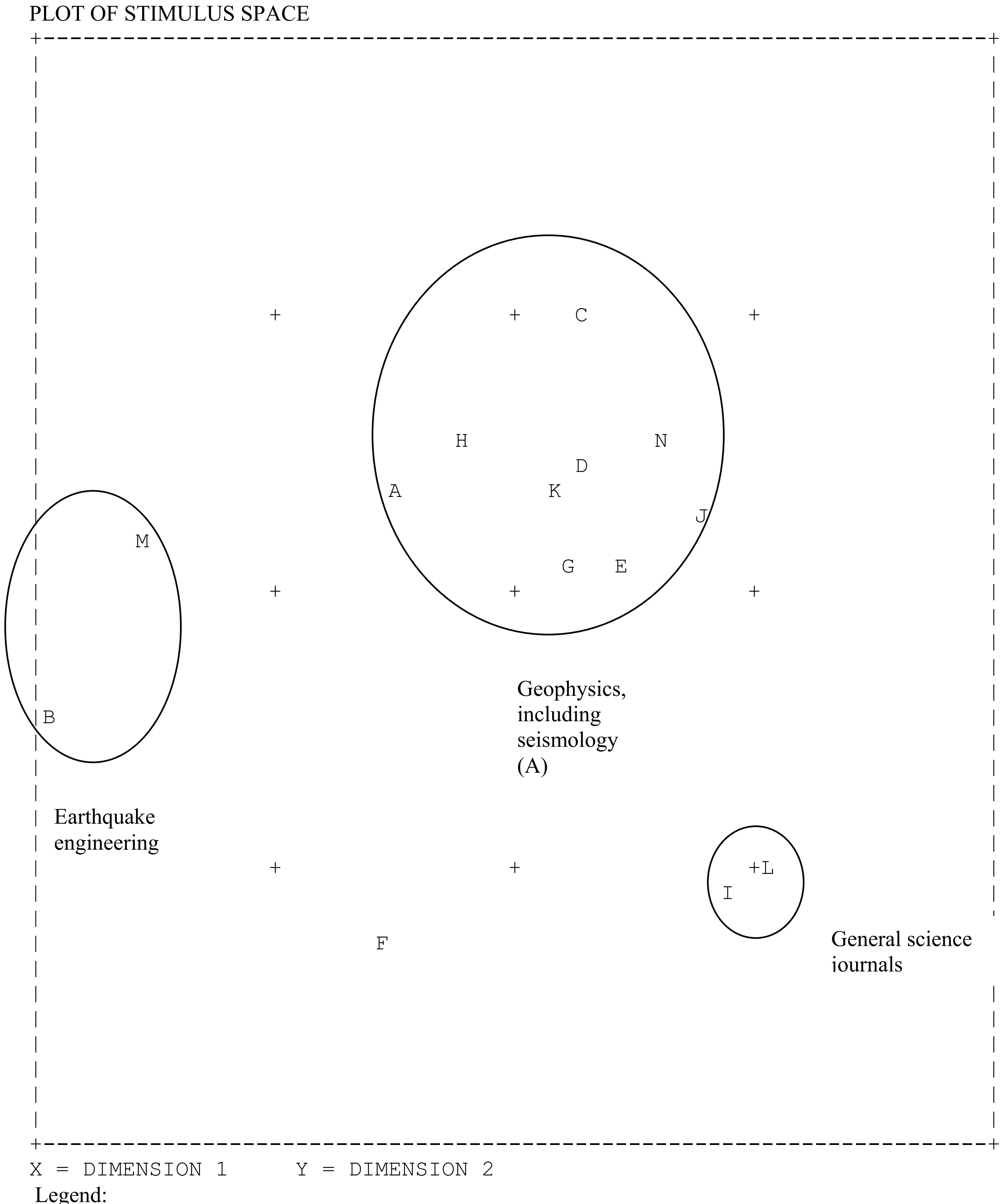

Legend:

| Indicating letter | **Earthquake Engineering Cluster Journal Names** |
|---|---|
| B | Earthquake Engineering and Structural Dynamics |
| M | Soil Dynamics and Earthquake Engineering |
| | |
| | **Geophysics Cluster Journal Names** |
| A | Bulletin of the Seismological Society of America |
| C | Fizika Zemli |
| D | Geophysics Journal International |
| E | Geophysics Research Letters |
| G | Journal of Geophysics Research |
| H | Journal of Physics of the Earth |
| J | Physics of Earth and Planetary Interiors |
| K | Pure and Applied Geophysics |
| N | Tectonophysics |
| | |
| | **General Science Cluster Journal names** |
| I | Nature |
| L | Science |
| | |
| | **Other Related but not Clustered Journals** |
| F | Geophysics |

## 3.1 Examining international linkages within the field

We expected to find that, because seismology is a dynamic, emerging field, one that relies on distributed knowledge and shared data, it is also more internationalised than older, related fields. This expectation arises from our thesis that emerging fields are also likely to have features reflecting a Mode 2 method of operating: they are interdisciplinary, team-based research. Thus, they are more likely to require a network of linkages in order to share and create knowledge. Moreover, because seismology requires access to unique, geographically-tied resources, we expected it to create the conditions where researchers meet, interact, and seek to conduct collaborative research on an international basis.

Contrary to our expectations, the initial review of the data does not show seismology as more heavily internationalised in terms of international co-authorship relations than geophysics in 2000. Table 2 displays the results of the bibliometric analysis of the articles published in 2000: within the geophysics cluster, internationally co-authored articles were 34 percent of all articles, while seismology shows 26.2 percent of articles as internationally co-authored for that year. As a percentage of articles published, geophysics is a more internationalised field than seismology.

| Year | Number of Articles Published | Number of Articles Internationally Co-authored | Percent of Articles Internationally Co-authored |
|---|---:|---:|---:|
| Geophysics | | | |
| 2000 | 2389 | 814 | 34.1 |
| Seismology | | | |
| 2000 | 409 | 107 | 26.2 |

**Table 2. A comparison of articles published in geophysics and seismology in 2000**

The social networks of the two fields also demonstrate a more densely interconnected network for geophysics than for seismology in 2000. Figures 5 and 6 illustrate the social networks that constitute geophysics (Fig. 5) and seismology (fig. 6) in 2000 when co-authorships are attributed to and counted at the country level. These different densities of the networks can be partly explained by the size of geophysics compared to seismology, but we will see below that after normalization for size effects using the cosine important differences will remain. The longer-established journals in the field of geophysics attract and publish many more articles per year than seismology journals, as can be noted from the data presented in Table 2.

The social network created by co-authorships in geophysics journals has a more “traditional” set of core countries than the ones exposed in the seismology network. Within geophysics, the core countries include those with the largest scientific enterprises, including the USA, Germany, England, Russia, France, the Netherlands, and Japan. Intriguingly, the network for seismology has a core that involves a number of non-traditional and developing countries. In fact, seismology’s core network does not include the countries named above. The core network consists of the western European countries of Italy, Germany, and Switzerland, and than a tight network of eastern European countries connecting to western Europe through Russia. Although the USA, France, and England publish more articles, they are not as tightly networked as the western and eastern European partners in the core represented in Figure 6.

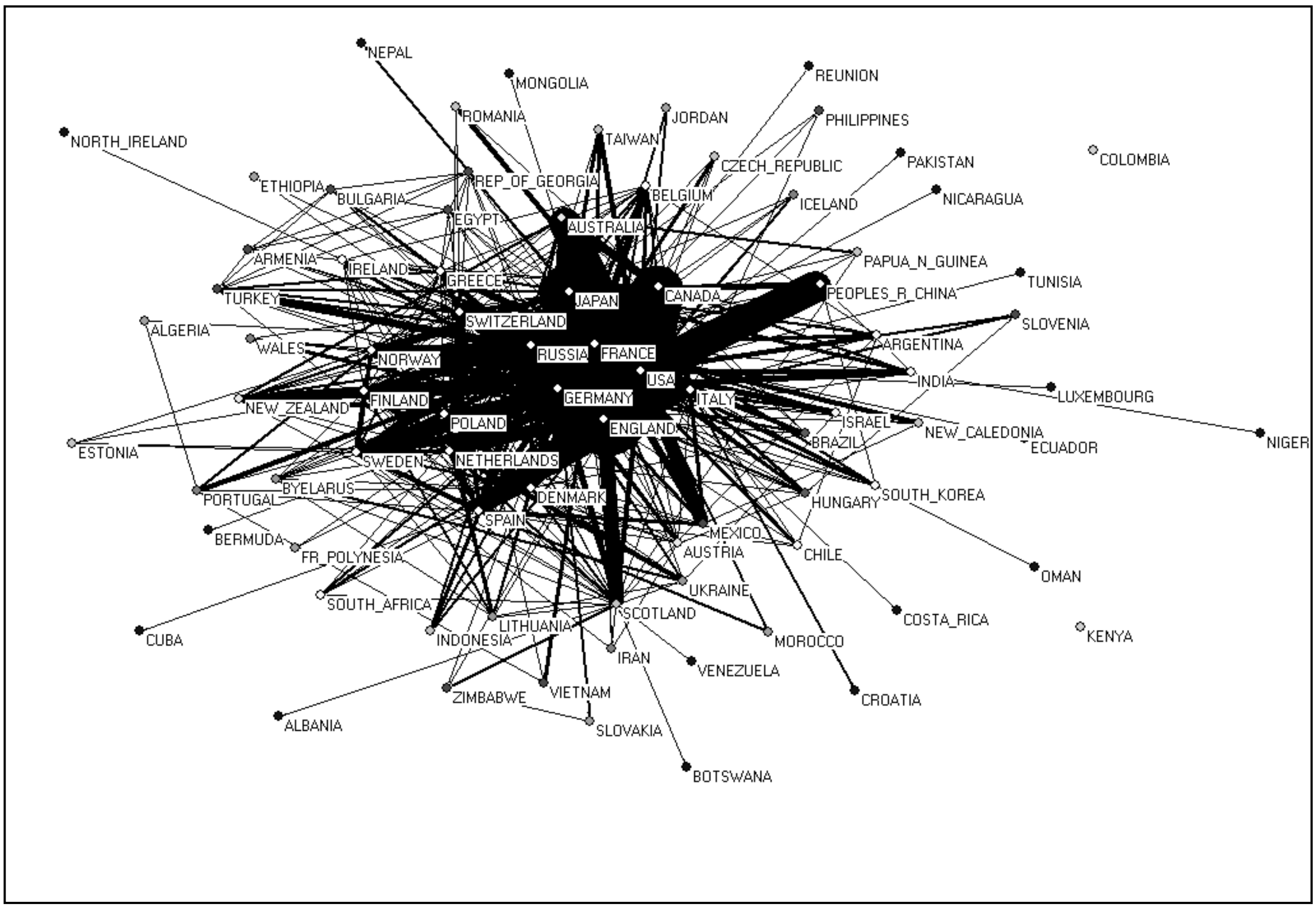

**Figure 5. The network of international co-authorships in geophysics, 2000**

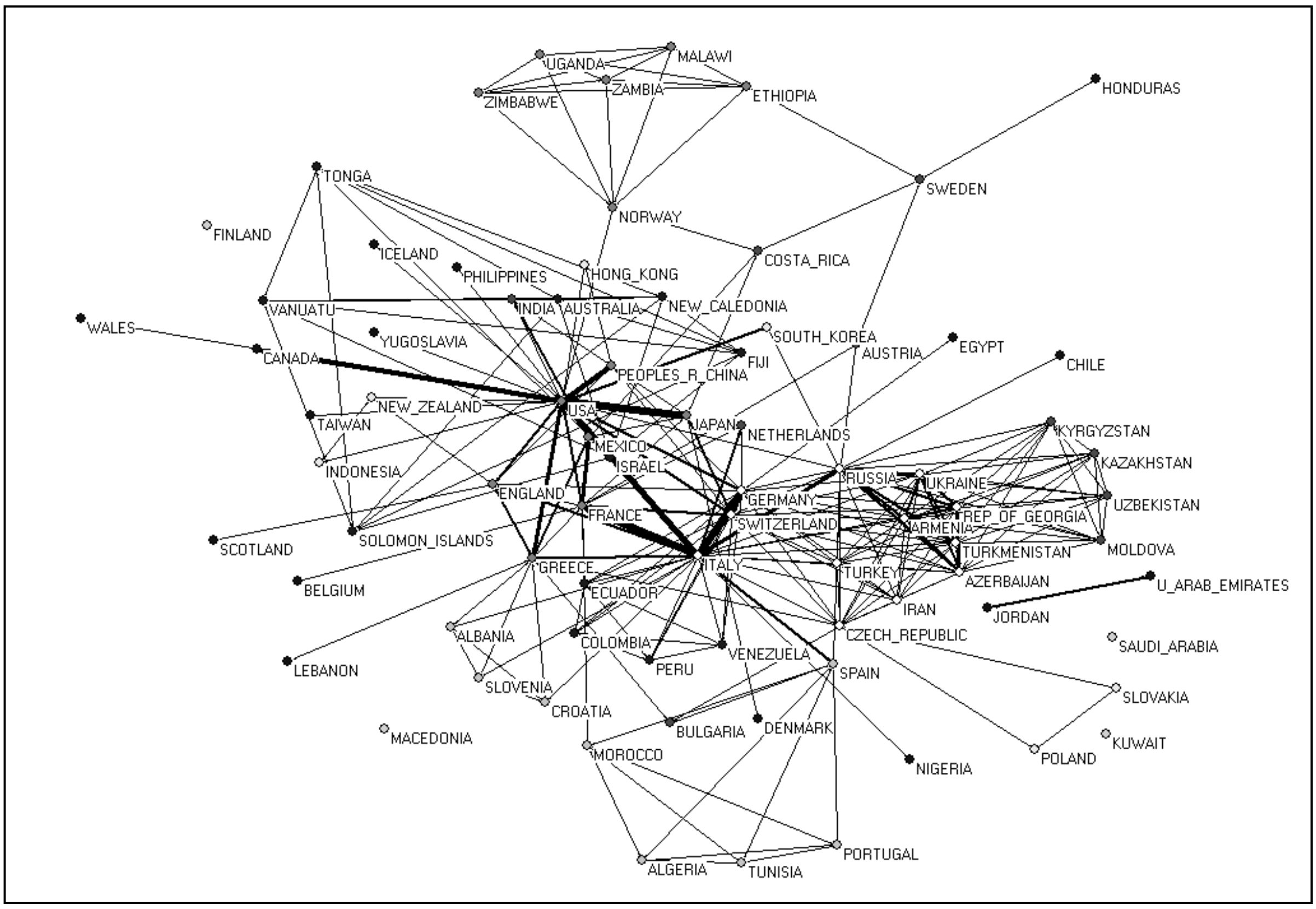

**Figure 6. The network of international co-authorships in seismology, 2000**

We tested the structure of the field further by conducting an analysis of the core of the structure of the co-authorship network of both fields. We found that geophysics also has a higher percentage of countries participating at the core of the network.

| Year | Number of Countries Participating in International Network-Geophysics | Number of Countries in the Core Group of International Co-author Networks |
|---|---|---|
| 2000 | 82 | 19 |
| 1990 | 56 | 9 |
| | Seismology | |
| 2000 | 80 | 12 |

**Table 3. Core group participation of geophysics and seismology, 2000**

We then sought to expose the organisation of the networks by conducting a cosine analysis to weight the data for the differences in size. Here we found significant differences between the two networks. Figures 7 and 8 reveal the hierarchy of the social networks in both geophysics and seismology for 2000 (the core countries are indicated by open circles). The number of countries in the core groups are similar in number, but the seismology network contains more tightly connected clusters with intense regional clusters when compared to geophysics. Italy seems to be the core country relating various networks. The geophysics map has a centre/periphery model with the USA, UK, France, and Germany in the very centre.

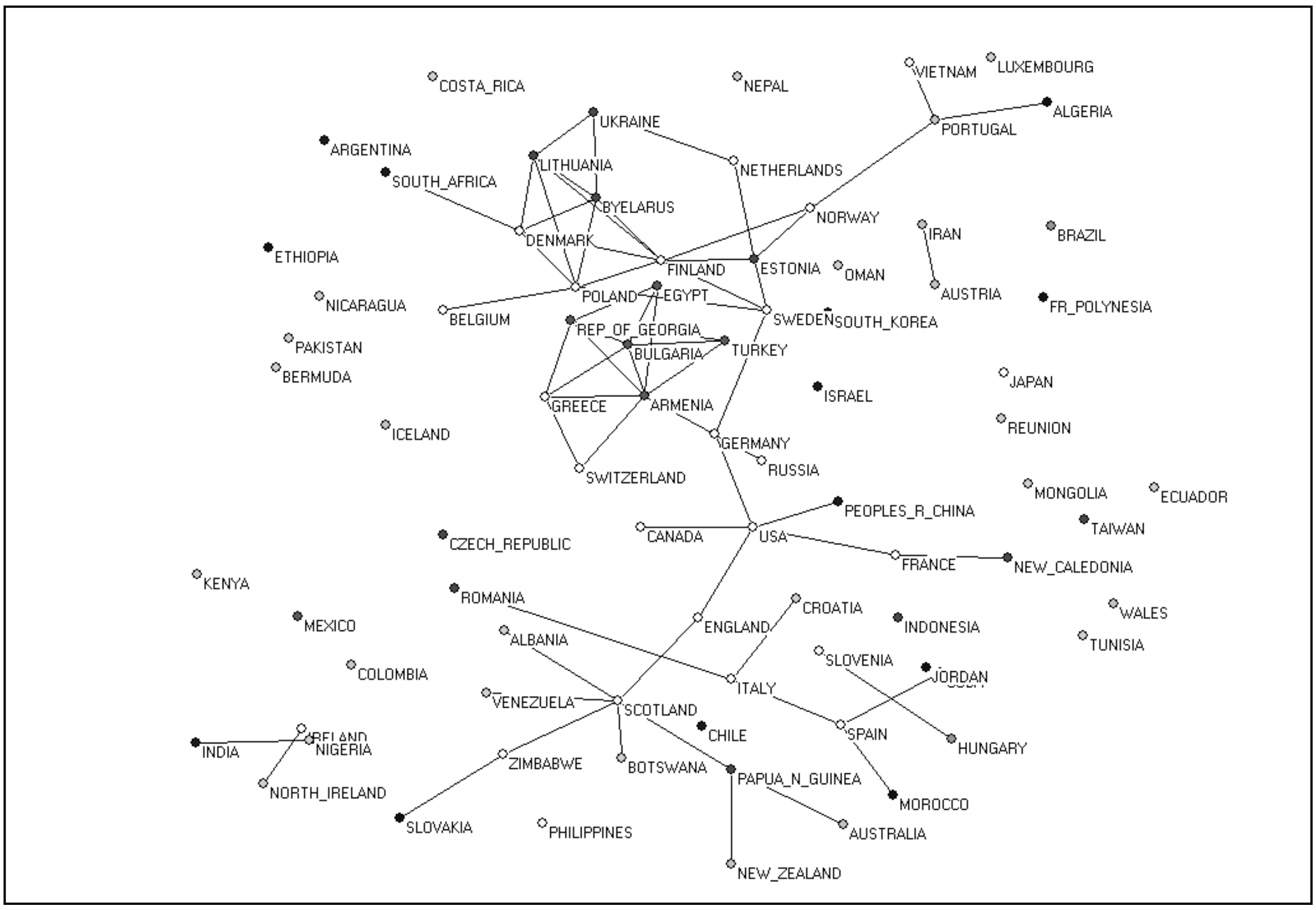

**Figure 7. Analysis of the network of geophysics collaboration at cosine greater than 0.1, 2000**

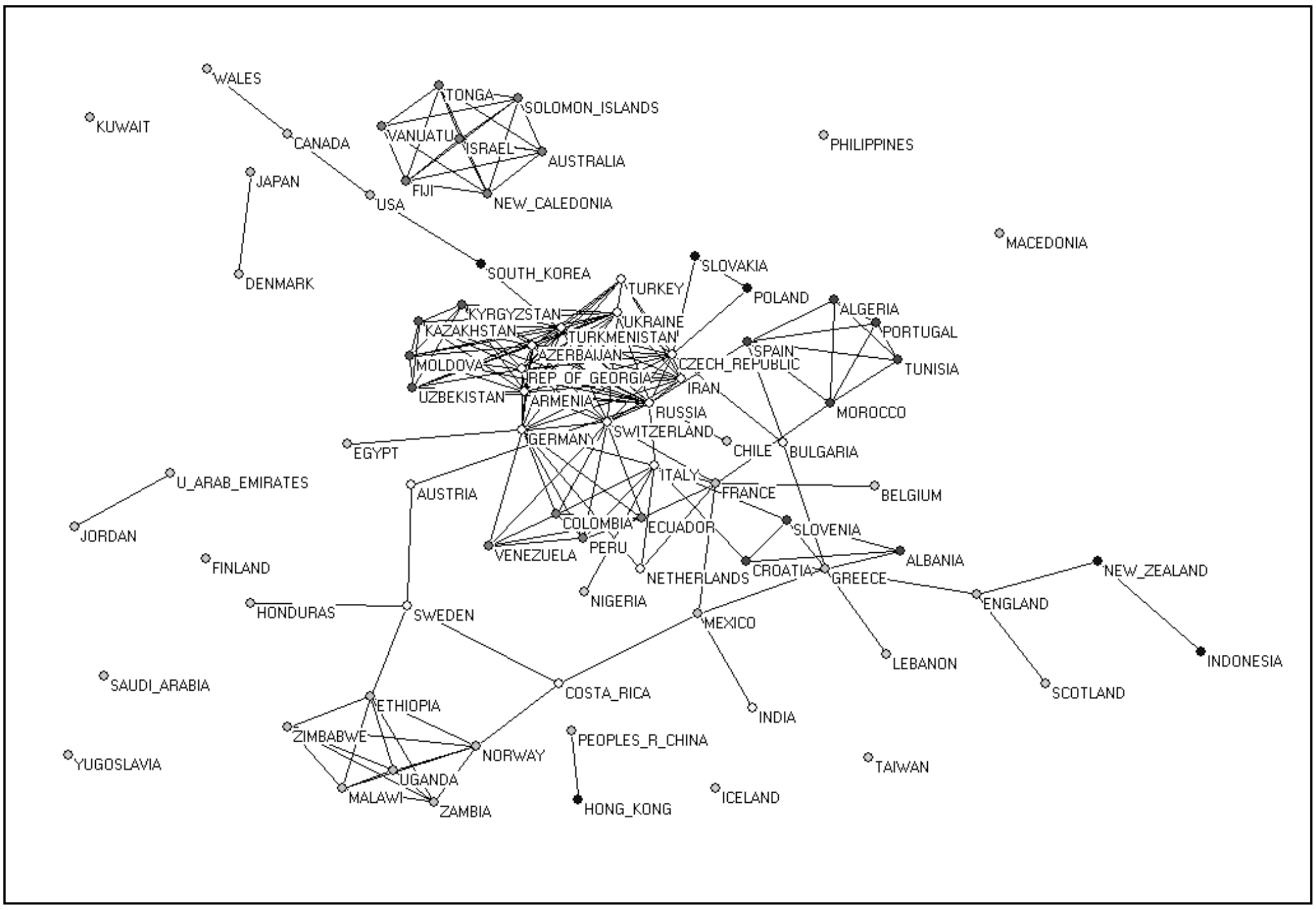

**Figure 8. Analysis of the network of seismology collaboration at cosine greater than 0.1, 2000**

These figures can both be compared to the cosine analysis of the network of the combined field of geophysics (encompassing seismology journals published at the time) as it existed in 1990, shown in Figure 9. The analysis suggests that the network of international collaboration in terms of co-authorship relations had very little structure in 1990. Local collaborations, for example with the Philippines, prevailed at that time.

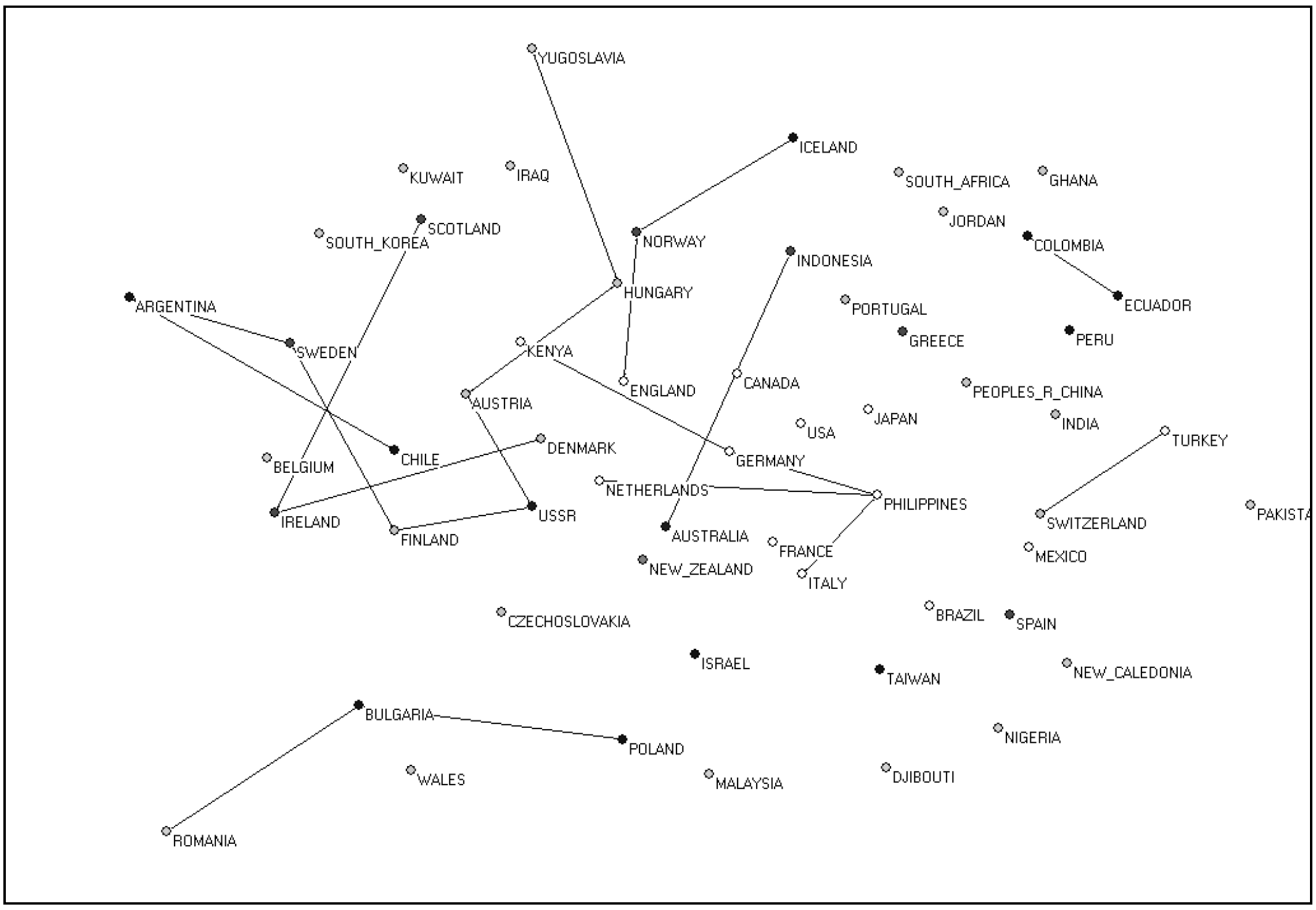

**Figure 9. Analysis of the network of collaborating countries at cosine greater than 0.1, geophysics (encompassing seismology), 1990**

# 4. Summary and conclusions

Seismology can be considered as an emerging field of science. This is demonstrated by the citing relationships showing that in the early 1990s, seismology did not exist within the citing environment as a field separate from geophysics. Contrary to our initial expectation, in 2000, the field of seismology does not have as high a percentage of internationally co-authored articles as its parent field, geophysics. An alternative explanation for this may be that as a new field emerges, it may have just a few centres of excellence where research is being conducted. As the research is recognized, practitioners from different countries seek to cooperate with the lead researchers, creating the international links over time. The “Mode 2” character may also lead to more grey literature than in an established field of science. The geographical shape of some of the clusters may reflect the influence of policy programs and their attempts to organize this field.

Although seismology did not contain as high a percentage of international articles as geophysics, we did have the unexpected finding that the core group of cooperating countries within seismology is a more distributed and broader network than that within geophysics. The networked group of seismology countries exposed in the cosine analysis has more members than would be expected given the structure of the parent field. The connections between these countries appear to be active collaborations. This suggests that emerging fields, although still growing their network of international connections, may do so by establishing a distributed set of initial connections throughout their social network rather than operating along a hub and spoke model.